\documentclass[prl,aps,twocolumn,groupedaddress,showpacs,floatfix,preprintnumbers,superscriptaddress]{revtex4-1}
\usepackage{amsmath,amssymb,multirow,bm}

\newcommand{\beq}{\begin{equation}}
\newcommand{\eeq}{\end{equation}}
\newcommand{\bea}{\begin{eqnarray}}
\newcommand{\eea}{\end{eqnarray}}
\usepackage{graphicx}
\preprint{}

\begin{document}

%\begin{frontmatter}

\title{Dynamical Coupling of Pygmy and Giant Resonances in Relativistic Coulomb Excitation}
\author{N.S.  Brady}
\email{nbrady@leomail.tamuc.edu}
\affiliation{Department of Physics and Astronomy, Texas A \& M University-Commerce, Commerce, Texas 75429, USA}
\author{T. Aumann}
\email{t.aumann@gsi.de} 
\affiliation{Institut f\"ur Kernphysik, Technische Universit\"at Darmstadt, Schlossgartenstr. 9, 64289 Darmstadt, Germany}
\affiliation{GSI Helmholtzzentrum f\"ur Schwerionenforschung, Planckstr. 1, 64291 Darmstadt, Germany}
\author{C.A. Bertulani}
\email{carlos.bertulani@tamuc.edu}
\affiliation{Department of Physics and Astronomy, Texas A \& M University-Commerce, Commerce, Texas 75429, USA}
\affiliation{Department of Physics and Astronomy,   Texas A \& M University, College Station, Texas 77843, USA}
\author{J.O. Thomas}
\email{jamesothomas@gmail.com}
\affiliation{Department of Physics and Astronomy, Texas A \& M University-Commerce, Commerce, Texas 75429, USA}

\date{\today}

\begin{abstract}

We study the Coulomb excitation of pygmy dipole resonances (PDR) in heavy ion reactions at 100 MeV/nucleon and above. The reactions $^{68}{\rm Ni}+^{197}{\rm Au}$ and  $^{68}{\rm Ni}+^{208}{\rm Pb}$ are taken as a practical examples. Our goal is to address the question of the influence of giant resonances on the PDR as the dynamics of the collision evolves. We show that the coupling to the giant resonances affects considerably the excitation probabilities of the PDR, a result that indicates the need of an improved theoretical treatment of the reaction dynamics at these bombarding energies.

\end{abstract}


\maketitle

%\section{Introduction}

The existence of collective vibrations in neutron-rich nuclei at low energies was suggested by Kubono, Nomura,  and collaborators in a 1987 proposal for the Japanese Hadron project which eventually became the J-PARC facility \cite{NS87}. This proposal was later given theoretical support by Ikeda \cite{Ike88} and collaborators. It took nearly two decades for experimental evidences of the existence of a collective low energy response to be found in neutron-rich nuclei, far from the valley of stability.  It is worthwhile mentioning that  direct breakup of light and loosely-bound projectiles, such as $^{11}$Be and $^{11}$Li were initially thought to be indicative of a collective nuclear response but it was shown to be a direct Coulomb dissociation of the weakly-bound  valence nucleons  \cite{BM90}.  Nowadays it is known as the Pygmy Dipole Resonances (PDR), which is the strength at low-lying energies due to the fragmentation of the nuclear response \cite{Lei01}. The energy spectrum is typically obtained with the experimental probe of choice, i.e., relativistic Coulomb excitation of projectiles  produced and accelerated in radioactive beam facilities  (for related reviews, see Refs. \cite{Aum05, AN13}).  In such a process,   the identification of pygmy resonances is done via their decay modes, usually via gamma or neutron emission, and the energy spectrum is obtained by invariant mass reconstruction from the energy of the fragments \cite{Lei01}.  PDRs are typically interpreted as due to the oscillation of the excess neutrons against a more tightly bound core.  

\begin{figure}
\begin{center}
\includegraphics[scale=0.4]{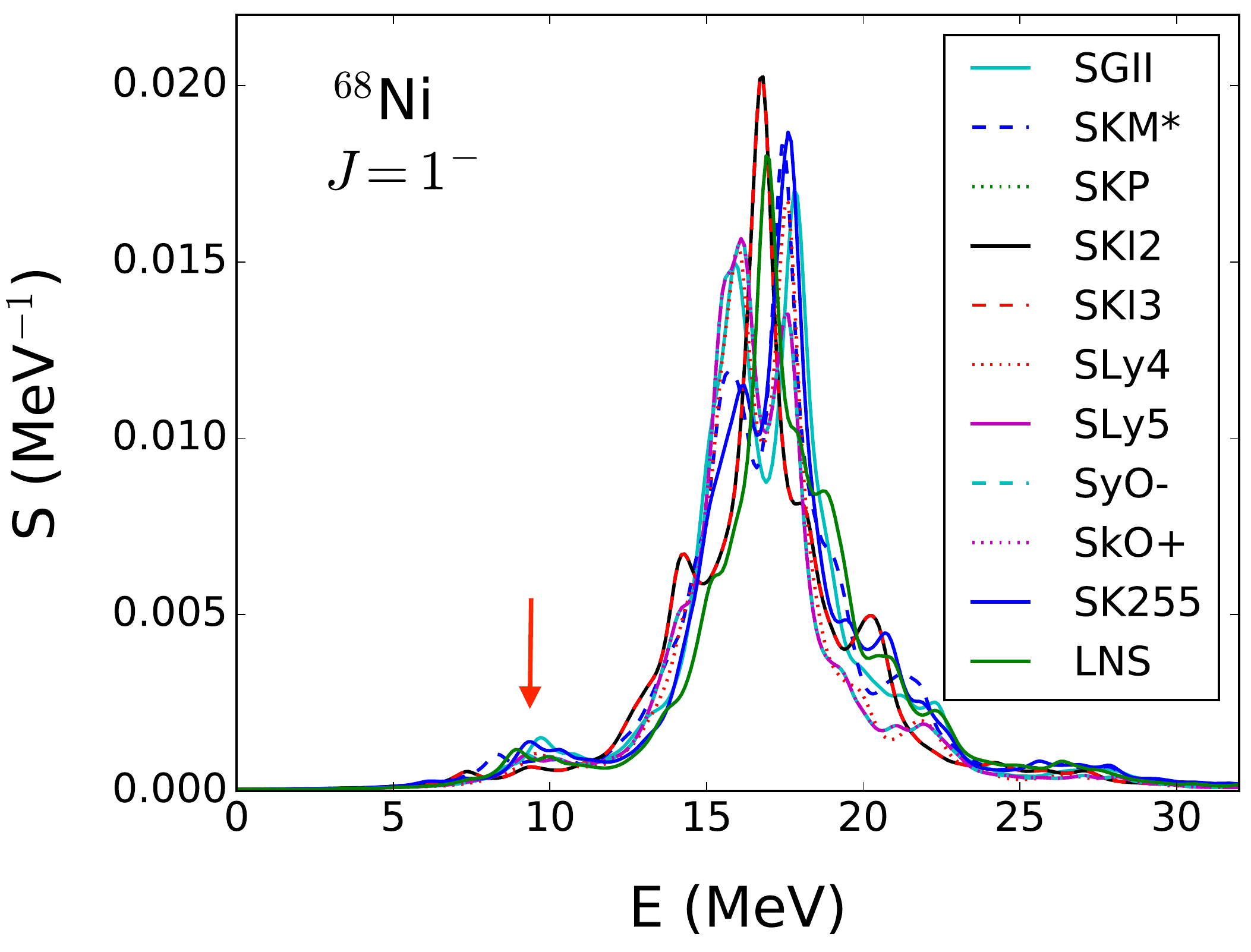}
\caption{Strength function for the E1 RPA response in $^{68}$Ni calculated with formalism described in Ref. \cite{Col13}.\label{figrpa}. The calculation is performed for several Skyrme interactions, shown in the figure inset. The arrow shows the location of the pygmy resonance.}
\end{center}
\end{figure}

Theories for giant resonances date back to when a simple hydrodynamical interpretation of protons  oscillating against the neutrons was used \cite{Gold48,Stein50}. Later on  microscopic calculations were developed based on the linear response theory \cite{Bar60}. Nowadays, an effort is being undertaken to describe nuclear collective motion with more elaborated models such as the time-dependent superfluid local density approximation \cite{Bul13,Ste15}.   Similarly, theoretical studies of the pygmy resonances have been developed based on the improvements of the hydrodynamical model \cite{Suz90,Vret01,Ber07}, and with microscopic theories such as the random phase approximation (RPA) and its variants \cite{Paa07,Kre09,Pon14,Pap14}. When reactions with radioactive beams became the focus of nuclear research in the last decades, it was soon realized that slight modifications of the linear response theory predict  a considerable concentration at low energies of the excitation strength in neutron-rich nuclei \cite{BF90,Ter91}.  As a word of caution, the amount of the sum rule exhausted by the nuclear response at low energies strongly depends on how the nuclear interaction, pairing, and other physical phenomena are incorporated in the theory \cite{Paa07,Kre09,Pon14,Pap14,BF90,Ter91}. As an example, the public code of Ref. \cite{Col13} has been used to calculate the E1 strength function, defined as 
\begin{equation}S(E)=\sum_\nu |\left< \nu||{\cal O}_L || 0\right>|^2 \delta(E-E_\nu), \label{E1str}\end{equation} 
defined for an RPA configuration space in terms of delta-function states $\nu$, where ${\cal O}_L$ is an electromagnetic operator.  A 1 MeV smearing of the fragmented strength function is introduced to yield a continuous distribution, shown in Figure \ref{figrpa} for the E1 response in $^{68}$Ni. In this case we used the option ${\cal O}_L = j_L(qr)$ in Eq. (\ref{E1str}), where $q=0.1$ fm$^{-1}$ was taken as representative of the momentum transfer. See  Ref. \cite{Col13} for more details. In this case, the strength function has dimensions of MeV$^{-1}$ and in the long-wavelength approximation, $qr\ll 1$, it is proportional to the usual response for electric multipole operators. The calculation is performed for several Skyrme interactions, as listed in the figure inset. The arrow shows the location of the expected pygmy dipole resonance. The results presented in the literature, e.g., in Refs. \cite{BF90,Ter91,Paa07,Kre09,Pon14,Pap14,BF90,Ter91} show a larger response in the PDR energy region due to adaptations in the model space and interactions. As of now, there is not a clear prediction of the precise location  of the pygmy strength. It could be in the range of  $7-12$ MeV for medium mass nuclei such as Ni isotopes. The amount of the sum rule exhausted by the pygmy resonance is also relatively unknown, although some models based on nuclear clusterization can yield up to a 10\% of the total strength \cite{AGB82}.   

One of the effects overseen in the experimental analysis of Coulomb excitation of pygmy resonances is the large excitation probability in Coulomb excitation at small impact parameters, leading to a strong coupling of  pygmy and giant resonances. This coupling is manifested in dynamical effects such as the modification of transition probabilities and cross sections for the excitation of the PDR. This has been observed in the past in the context of the excitation of double giant dipole resonances (DGDR) \cite{BB88,ABE98,BP99}. The observation of the DGDR in experiments is a consequence of higher-order effects in relativistic Coulomb excitation and arises because the large excitation probabilities of giant resonances in heavy ion collisions at small impact parameters. The dynamical coupling between the usual giant resonances and the DGDR is very strong, as shown, e.g., in Ref. \cite{Bert96}.   In the present work we make an assessment of this effect on the excitation of the PDR using the relativistic coupled channels (RCC) equations introduced in Ref. \cite{Ber05}.  

\begin{figure}
\begin{center}
\includegraphics[scale=0.34]{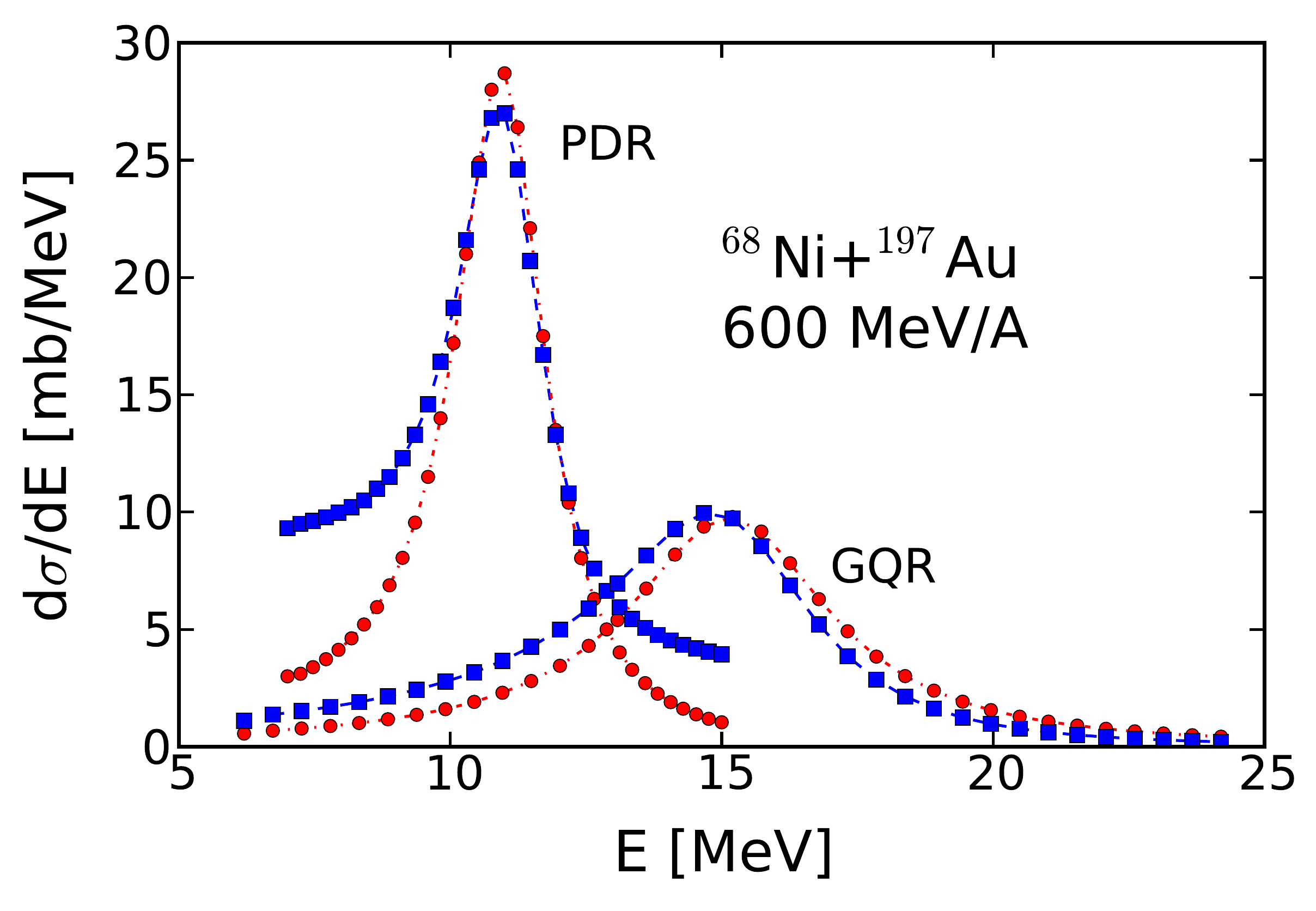}
\caption{Coulomb excitation cross section as a function of the excitation energy of 600 MeV/nucleon  $^{68}$Ni projectiles incident on $^{197}$Au targets. The filled circles represent the calculations using first-order perturbation theory, while the filled squares are the results of coupled-channel calculations.}
\label{cc1}
\end{center}
\end{figure} 

The S-matrix, $S_\alpha \left(z, b\right)$, for Coulomb excitation is obtained from the RCC equations \cite{Ber05}
\begin{equation}
iv {\partial S_\alpha (z, b) \over \partial z} = \sum_{\alpha'} \left< \alpha \left| {\cal M}_{EL} \right| \alpha' \right> S_{\alpha '}(z,b) e^{- i{(E_{\alpha '} -E_\alpha)z/ \hbar v}},\label{rcc}
\end{equation}
where $v$ is the projectile velocity and ${\cal M}_{EL}$ is the electromagnetic operator for electric dipole (E1) and quadrupole (E2) transitions connecting states $\alpha$ and $\alpha'$ satisfying the selection rules of their intrinsic angular momenta and parities.  The ground state is denoted by $\left|0\right>=\left| E_0 J_0 M_0\right>$ and the excited states by $\left|\alpha \right>=\left| E_\alpha J_\alpha M_\alpha\right>$, where $\left|EJM\right>$ labels intrinsic energy and angular momentum quantum numbers. In the long-wavelength approximation the electromagnetic operators are given by \cite{Ber05}
\begin{equation}
{\cal M}_{{\rm E1m}}=\sqrt{\frac{2\pi}{3}} \xi Y_{1m}(  \mathbf{\hat
{\mbox{\boldmath$\xi$}}})  \frac{\gamma Z_{\rm T}e^2}{\left(
b^{2}+\gamma ^{2}z^{2}\right)^{3/2}}\left\{
\begin{array}
[c]{c}%
\mp b\ \ (\mathrm{if}\ \  m=\pm1)\\
\sqrt{2}z\ \ (\mathrm{if}\ \  m=0)\
\end{array}
\right.  \label{eq6}
\end{equation}
where $\mbox{\boldmath$\xi$}$ is the intrinsic coordinate of the excited nucleus and $Ze$ is the charge of the nucleus giving rise to the electromagnetic field (in our case, the target). For E2 transitions the electromagnetic operator is \cite{Ber05}
\begin{align}
{\cal M}_{{\rm E2}\mu}= &  \sqrt{\frac{3\pi}{10}}\xi^{2}Y_{2\mu}(
\mathbf{\hat {\mbox{\boldmath$\xi$}}})  \frac{\gamma
Z_{\rm T}e^2}{\left(  b^{2}+\gamma
^{2}z^{2}\right)  ^{5/2}}\nonumber\\
&  \times\left\{
\begin{array}
[c]{c}%
b^{2}\ \ \ \ (\mathrm{if}\ \ \ \mu=\pm2)\\
\mp2\gamma^{2}bz\ \ \ \ (\mathrm{if}\ \ \ \mu=\pm1)\\
\sqrt{2/3}\left(  2\gamma^{2}z^{2}-b^{2}\right)  \ \ \ \ (\mathrm{if}%
\ \ \ \mu=0)\, .
\end{array}
\right.  \label{eq7}%
\end{align}
Note that ${\cal M}_{ELm} = f_{ELm}({\bf r}) {\cal O}_{ELm}$, where ${\cal O}_{ELm}=\xi^LY_{Lm}(\mathbf{\hat {\mbox{\boldmath$\xi$}}})$ is the usual electric operator, and $ f_{ELm}({\bf r}) $ is a function of the projectile-target relative position ${\bf r}=({\bf b},z)$.

The coupled equations \eqref{rcc} are solved by using $S_\alpha (z\rightarrow -\infty) = \delta_{\alpha 0}$. For high energies and very forward angles, the cross sections for the $\left|0\right>\longrightarrow \left| \alpha \right>$ transition is given by
\begin{equation}
{d\sigma_\alpha\over dE} = 2 \pi w_\alpha (E) \int db \ b \exp\left[-2\chi(b)\right] \left|S_\alpha \left(z\rightarrow \infty, b\right)\right|^2 , \label{dsde}
\end{equation}
where $w_\alpha(E)$ is the density of final states, $b$ is the impact parameter in the collision, and $\chi(b)$ is the eikonal absorption phase given by
\begin{equation}
\chi(b)={\sigma_{NN}\over 4\pi}\int dq \ q \rho_1(q) \rho_2(q) J_0(qb),
\end{equation}
where $\sigma_{NN}$ is the experimental value of the total nucleon-nucleon cross section with medium corrections added according to Refs. \cite{BC10,Kar13} and $\rho_i(q)$ is the Fourier transform of the  ground state densities of the nuclei obtained from fitting to electron scattering experiments \cite{Vri87} for $^{197}$Au and using Hartree-Fock-Bogoliubov calculations for $^{68}$Ni with the SLy4 interaction. It is worth mentioning that the use of the eikonal absorption phase to cut down the Coulomb excitation mechanism at small impact parameters has been introduced in Ref. \cite{ABS95} for the first time to calculate cross sections relevant to GDR and DGDR excitations. The effects of nuclear excitation  have been subtracted in the experiments \cite{Wie09,Ros13,Sav13}. Therefore, we did not include nuclear excitations, and possible interferences, in these calculations. 

We consider the excitation of  $^{68}$Ni on $^{197}$Au and $^{208}$Pb targets at  600 and 503 MeV/nucleon, respectively. These reactions have been experimentally investigated in Refs. \cite{Wie09,Ros13}. In the first experiment a pygmy dipole resonance in $^{68}$Ni was identified at $E_{PDR} \simeq 11$ MeV with a width of $\Gamma_{PDR} \simeq 1$ MeV, exhausting about 5\% of the Thomas-Reiche-Kuhn (TRK) energy-weighted sum rule. The identification was done with the analysis the excitation and decay via gamma emission. In the second experiment  the PDR centroid energy was found to be at 9.55 MeV, with a 2.8\% fraction of the TRK sum rule, a width of 0.5 MeV, and the PDR identification was done by measuring the neutron decay channel of the PDR. In this work our objective is to study the effects of the coupling between the several giant resonances with the PDR, and therefore we will only calculate the excitation function d$\sigma$/dE without concern for the decay channels. 

One needs a model for bound and continuum discretized wavefunctions  entering the matrix elements $\left< \alpha \left| {\cal M}_{EL} \right| \alpha' \right>$ in Eq.  \eqref{rcc}. The wavefunctions can also be used to calculate the response functions, $dB_{EL}/dE= \sum_{spins}w_{\alpha'}\left|\left< \alpha || {\cal O}_{EL} || \alpha' \right>\right|^2$,  with an appropriate sum over angular momentum coefficients. Instead of considering  numerous nuclear structure  models for the wavefunctions and to dwell with the microscopic properties of the pygmy and giant resonances, we assumed Lorentzian forms for the response functions  $dB_{EL}/dE$ with a given fraction of the sum-rule, and discretized them in energy bins to obtain the reduced matrix elements $\left|\left< \alpha || {\cal M}_{EL} || \alpha' \right>\right|^2 \propto \Delta E_x (dB_{EL}/dE)|_{E=E_x}$, where $E_x=E_{\alpha '} - E_\alpha$. A phase convention can be found so that the reduced matrix elements are real. They are then used to deduce the matrix elements $\left< \alpha | {\cal M}_{EL} | \alpha' \right>$ in Eq. \eqref{rcc}  with proper care of the corresponding angular momentum coefficients (see, e.g., Ref. \cite{BP99}).

The Lorentzian functions are centered at the energies  $E_{PDR}$ for pygmy dipole resonances and $E_{GDR}$ ($E_{GQR}$) for the isovector (isoscalar) giant dipole (quadrupole)  resonances. Their respective widths are denoted by $\Gamma_{PDR}$, $\Gamma_{GDR}$ and $\Gamma_{GQR}$. Further, this strength function is subdivided into 35 energy bins  bins centered around the PDR energy and  the same number of energy bins centered around the GDR and GQR resonances.   We use $E_{PDR}=11$ MeV, consistent with Refs. \cite{Wie09,Ros13}, but a full width at half maximum of 2 MeV, which is more in line with theoretical calculations \cite{Vret01,Ber07,Paa07,Kre09,Pon14,Pap14,BF90,Ter91}  than with the experimental data \cite{Wie09,Ros13}. The larger PDR width also allows us to better determine the higher-order effects on the modification of the tails of the PDR.  For the (isovector) 1$^-$ giant dipole resonance (GDR) we assume $E_{GDR}= 17.2$ MeV and $\Gamma_{GDR}=4.5$ MeV and for the (isoscalar) 2$^+$ giant quadrupole resonance (GQR) we take $E_{GQR}=15.2$ MeV and $\Gamma_{GQR}=4.5$ MeV. The centroid and width for the GQR are not experimental, but approximate estimates based on  the  systematics of GQR excitation in nickel isotopes \cite{Youn92}. The total number of channels are $35\times 3 + 35\times 5 + 35 \times 5 +1 = 456$ including all magnetic substates and the ground state, assumed to be a $0^+$ state. The calculations are CPU intensive but can be reduced for the practical purposes because the major dynamical effect arises from the coupling of the PDR with the GQR via the dominant E1 interaction at relativistic bombarding energies.  The number of channels can also be reduced by a factor of 2 by means of a coarser binning of the PDR and GQR states with a loss of accuracy at the level of 10\%. With the number of channels mentioned earlier our calculations converge to within 1\%.  

In Figure \ref{cc1} we show the first-order Coulomb excitation cross sections  of the PDR and GQR separately, as a function of the excitation energy and for 600 MeV/nucleon  $^{68}$Ni projectiles incident on $^{197}$Au targets. The filled circles represent the calculations using first-order perturbation theory, while the filled squares are the results of coupled-channel calculations. As shown in Ref. \cite{BB88}, first-order Coulomb excitation cross sections can be obtained by means of the relation 
\begin{equation}
{d\sigma\over dE} = \sum_{\pi L} {N_{\pi L} (E)\over E} \sigma^{(\gamma)}_{\pi L} (E), 
\end{equation}
where $\pi L$ denotes the multipolarity, $N_{\pi L}$ are the virtual photon numbers, and $\sigma^{(\gamma)}_{\pi L}$ are the cross sections for real photons with multipolarity $\pi L$. The virtual photon numbers include the same absorption coefficient as in Eq. \eqref{dsde} \cite{ABS95}. The sum runs over the relevant multipoles, here E1 and E2 stand for $1^-$ and $2^+$ excitations, respectively. It is quite evident from the figure that the coupling between these states has a visible impact on the energy dependence of the cross sections. We notice that according to the Brink-Axel hypothesis, a giant resonance can be excited on top of any other state in a nucleus \cite{Bri55,Axe62}. Therefore, the couplings are in fact a manifestation of (PDR$\otimes$GQR)$_{1^-}$, (PDR$\otimes$PDR)$_{2^+}$, (PDR$\otimes$GDR)$_{2^+}$ and (GDR$\otimes$GQR)$_{1^-}$ states which are of our interest, as they build up components of the PDR, GDR and GQR. The importance of our findings lies on the reliability of the  experimental extraction of the PDR strength relative to that of the GDR. 

\begin{figure}[t]
\begin{center}
\includegraphics[scale=0.47]{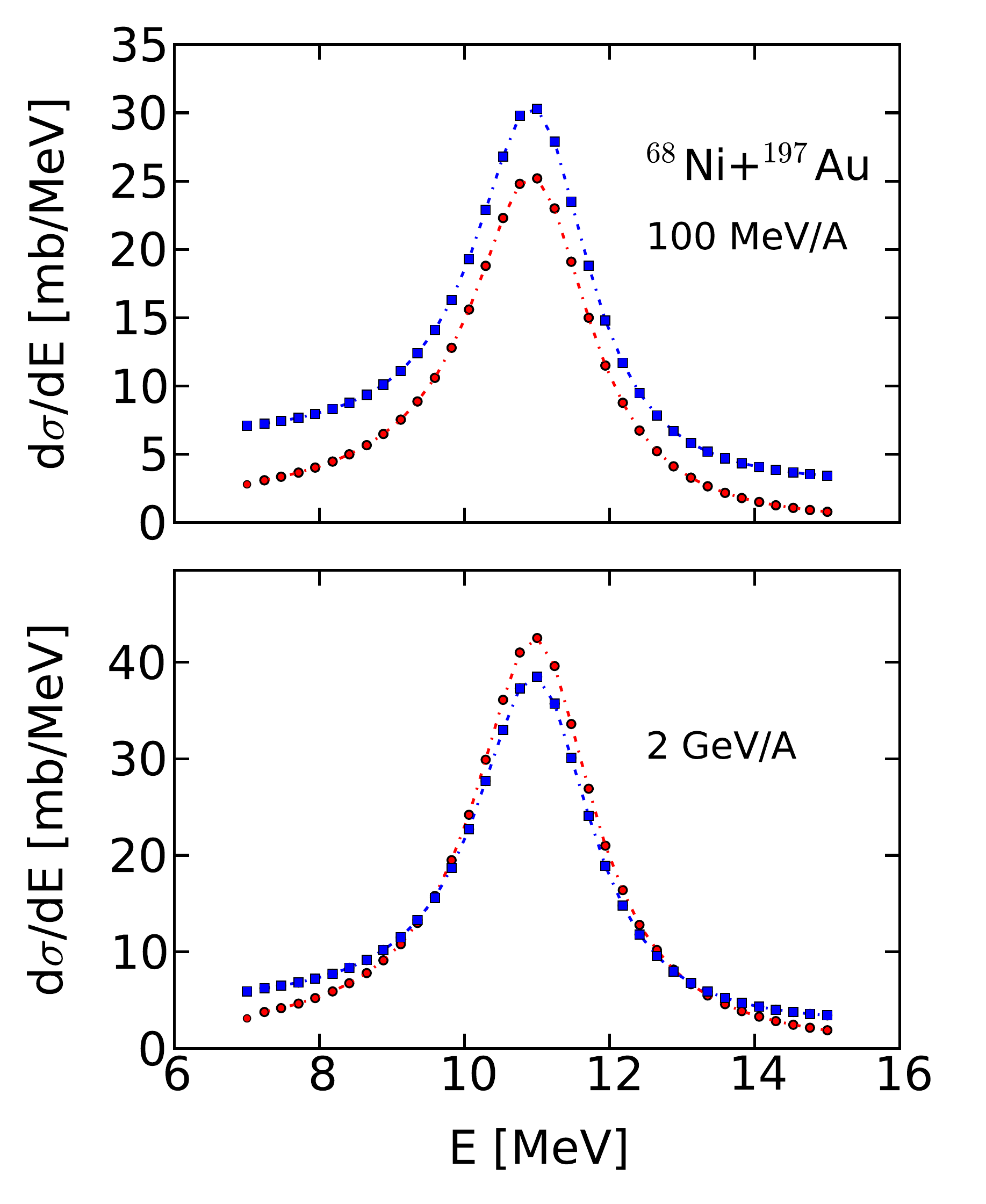}
\caption{Coulomb excitation cross section as a function of the excitation energy of  $^{68}$Ni projectiles incident on $^{197}$Au targets at two laboratory energies. The filled circles represent the calculations using first-order perturbation theory, while the filled squares are the results of coupled-channel calculations.}
\label{cc2}
\end{center}
\end{figure} 

The dynamical calculations show that not only the strength, but also the width of the PDR is modified appreciably  due to the coupling to the GQR. In Figure  \ref{cc1}  the modification of the population of PDR and GQR states are shown separately.  Because the $1^-$ states in the GDR region  are very weakly affected by the coupling to the other states, we left them out of the figure as we want to concentrate on the PDR excitation spectrum.  The GDR excitation dominates  in  the high energy region by a factor of $2-3$ times that of the 2$^+$ states. The main modifications in the excitation spectrum comes from the couplings ${\rm PDR} \longleftrightarrow {\rm GQR} \longleftrightarrow {\rm PDR}$ by E1 fields, while the couplings ${\rm PDR}  \longleftrightarrow {\rm GDR}  \longleftrightarrow  {\rm PDR}$ by E2 fields contribute very little to the 1$^-$ states in the PDR energy region. We also see in Figure \ref{cc1} that the tails of the PDR, and to a minor extent those of the GQR, are appreciably modified. A small shift of the peaks also occurs, although barely visible for the PDR, it is evident for the GQR. One has to keep in mind that the low energy tail of the GDR will modify the strength and shape of the PDR. For the case we consider here, a GDR strength of the order of 3.8\% lies in the region of the PDR and the PDR shape with only slightly be influenced by this low energy tail. However, these effects have been considered in the experimental analyses \cite{Wie09,Ros13}. In this work we are interested in the higher-order effects which have been so far ignored.  

In Figure \ref{cc2} we singled out the energy region of the pygmy resonance and we plot the results of our calculations for two different bombarding energies: 100 MeV/nucleon and 2 GeV/nucleon, with the same notation as in Figure \ref{cc1}. The coupling effects change dramatically. At the lower energy the influence of the giant resonances is to increase appreciably the response in the energy region of the PDR, while at the higher energy the effect of coupling is much smaller and the tendency is to slightly decrease the PDR excitation cross section. This result is expected because at  energies around 100 MeV/nucleon the E2 field is dominant, with an appreciable increase of  the excitation of the GQR and a consequently strong feedback to the PDR via subsequent E1 transitions.

\begin{figure}[t]
\begin{center}
\includegraphics[scale=0.38]{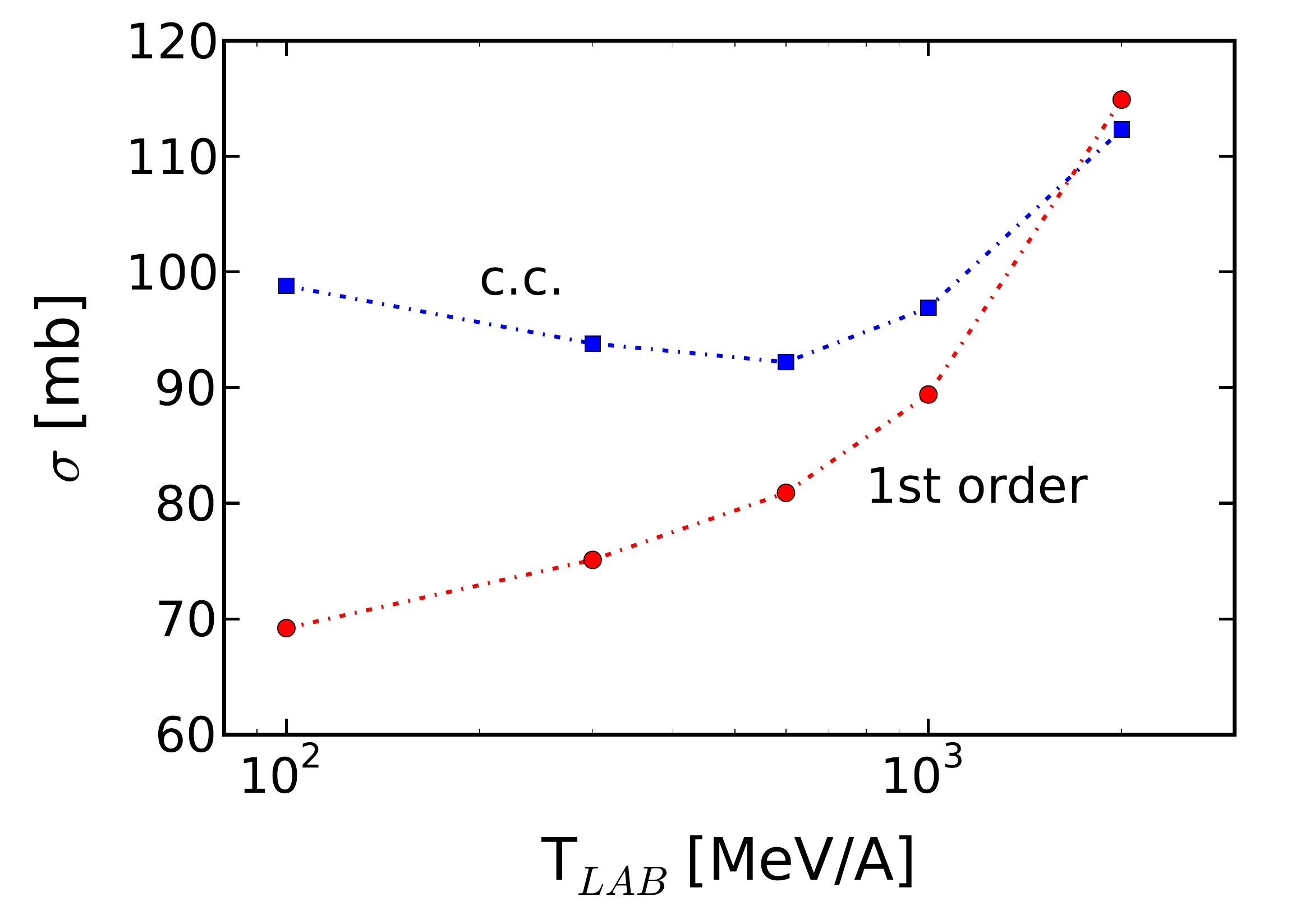}
\caption{Coulomb excitation cross sections of the PDR as a function of the bombarding energy of   $^{68}$Ni projectiles incident on $^{197}$Au targets. The filled circles represent the calculations using first-order perturbation theory, while the filled squares are the results of coupled-channel calculations.}
\label{CS}
\end{center}
\end{figure} 

In Figure \ref{CS} we show the Coulomb excitation cross sections of the PDR  as a function of the bombarding energy of  $^{68}$Ni projectiles incident on $^{197}$Au targets. The filled circles represent the calculations using first-order perturbation theory, while the filled squares are the results of coupled-channel calculations. At lower energies the deviation is clearly more pronounced. At 600 MeV/nucleon the cross section for excitation of the PDR changes from 80.9 mb obtained with the virtual photon method to 92.2 mb with the coupled-channels calculation. If this is reflected in the extracted PDR strength from the experimental data, it amounts to an appreciable change of 14\%. It implies a reduction by approximately the same amount of the strength needed to reproduce the experimental data.  

We have also performed calculations for  $^{68}{\rm Ni}+^{208}{\rm Pb}$ at 503 MeV/nucleon, corresponding to the experiment of Ref. \cite{Ros13}. The Coulomb excitation cross section for the PDR in $^{68}$Ni to first-order is  found to be 58.3 mb, while when couplings to the giant resonances are included, the cross section increases to 71.2 mb, i.e., an important 18.1\% correction. The dipole polarizability is defined by
\begin{equation}
\alpha_D = {\hbar c\over 2 \pi^2} \int dE {\sigma(E) \over E^2},
\end{equation}
where $\sigma(E)$ is the photo-absorption cross section. The value of $\alpha_D$ extracted from the experiment in Ref. \cite{Ros13} is $3.40$ fm$^3$ while to reproduce the experimental cross section with our dynamical calculations we have $\alpha_D = 3.16$ fm$^3$, a small but non-negligible correction. If a linear relationship between the dipole polarizability and the neutron skin is assumed \cite{Pie11},  a reduction of the neutron skin from 0.17 fm, as reported in Ref. \cite{Ros13}, to 0.16 fm is to be expected. Such a correction still lies within the experimental uncertainty of 7\% for $\alpha_D$ and 0.02 fm for the neutron skin \cite{Ros13}. However, coupling effects should be taken into consideration in the future when more precise data will become available, in particular, if the measurement is performed at lower bombarding energies.

We conclude that due to the large Coulomb excitation probabilities of giant resonances in heavy ion collisions at energies around and above 100 MeV/nucleon,  the excitation of the PDR is also appreciably modified due to the coupling between the $1^-$ and $2^+$ states. Our calculations are simplified with the use of a Lorentz-like distribution of the electromagnetic response and sum-rules, without a detailed nuclear structure model. In the future it might be possible to carry out nearly ``ab-initio" calculations based on a microscopic theory, coupled with a proper reaction mechanism. A known alternative, already used in previous studies of multiphonon  resonances \cite{PB98}, is to  use individual states calculated with the RPA or other microscopic models together with higher order perturbation theory. Finally,  one might also use advanced mean-field time-dependent method such as that developed in Ref. \cite{Ste15}. The relevance to derive rather accurate dipole strength distributions from electromagnetic  excitation of the PDR is mainly due to the extraction of the dipole polarizability \cite{Ros13}, which is an important observable to constrain the symmetry energy, and is thus also important for the understanding of neutron-star properties. The low-energy response is particularly important for the polarizability due to the inverse weighting with energy. This opens really exciting possibilities for the studies of the pygmy resonance in nuclei and use it as a tool for applications in nuclear astrophysics.

\section*{Acknowledgement} 

This work was supported in part by the U.S. DOE grants DE-FG02-08ER41533 and  the U.S. NSF Grant No. 1415656. We thank HIC for FAIR for supporting visits of one of us (C.A.B.) at the Technische Universit\"at Darmstadt. This project was also supported (T.A.) by BMBF, and GSI Darmstadt.


\begin{thebibliography}{99}
\bibitem{NS87} T. Nomura and S. Kubono, ``Soft giant resonance", Experimental proposal to the Japanese Hadron project (now J-PARC), June 1987.
\bibitem{Ike88} K. Ikeda, INS Report JHP-7 (1988) (in Japanese); K. Ikeda. Nucl. Phys. A 538, 355c (1992).
\bibitem{BM90} C.A. Bertulani and M.S. Hussein, Phys. Rev. Lett. 64, 1099 (1990).
\bibitem{Lei01} A. Leistenschneider et al., Phys. Rev. Lett. 86, 5442 (2001).
\bibitem{Aum05} T. Aumann, Phys, Eur. J A 26, 441 (2005).
\bibitem{AN13} T Aumann and T Nakamura, Phys. Scr. T152, 014012 (2013).
\bibitem{Gold48} M. Goldhaber and E. Teller, Phys. Rev. 74, 1046 (1948).
\bibitem{Stein50} H. Steinwedel and J. H. D. Jensen, Z. Naturforschung 5A, 413 (1950).
\bibitem{Bar60} M. Baranger, Phys. Rev 120, 957 (1960).
\bibitem{Bul13} A. Bulgac, Annu. Rev. Nucl. Part. Sci. 63, 97 (2013).
\bibitem{Ste15} I. Stetcu, et al., Phys. Rev. Lett. 114, 012701 (2015).
\bibitem{Suz90} Y. Suzuki, K. Ikeda, and H. Sato, Prog. Theor. Phys. 83, 180 (1990).
\bibitem{Vret01} D. Vretenar et al., Nucl. Phys. A 692, 496 (2001).
\bibitem{Ber07} C.A. Bertulani, Phys. Rev. C 75, 024606 (2007); Nucl. Phys. A 788, 366 (2007).
\bibitem{Paa07} N. Paar, D. Vretenar, E. Khan, and G. Col\`o,  Rept. Prog. Phys. 70, 691 (2007).
\bibitem{Kre09} S. Krewald and J. Speth,  Int. J. Mod. Phys. E18, 1425 (2009).
\bibitem{Pon14} V.Yu. Ponomarev, J. Phys: Conf. Series 533, 012028 (2014).
\bibitem{Pap14} P. Papakonstantinou, H. Hergert, V.Yu Ponomarev, and R. Roth,  Phys. Rev. C 89, 034306 (2014).
\bibitem{BF90} G. Bertsch and J. Foxwell, Phys. Rev. C41 (1990) 1300. (Erratum: Phys. Rev. C42, 1159 (1990).
\bibitem{Ter91} N. Teruya, C.A. Bertulani, S. Krewald, H. Dias and M. S. Hussein, Phys. Rev. C 43, 2049 (1991).
\bibitem{Col13} G. Col\`o, L. Cao, N. V. Giai and L. Capelli, Comput. Phys. Comm. 184, 142 (2013).
\bibitem{AGB82} Y. Alhassid, M. Gai, and G.F. Bertsch, Phys. Rev. Lett. 49,1482 (1982).
\bibitem{BB88} C.A. Bertulani and G. Baur, Phys. Reports 163, 299 (1988).
\bibitem{ABE98} T. Aumann, P. F. Bortignon, and H. Emling, Annu. Rev. Nucl. Part. Sci. 48, 351 (1998).
\bibitem{BP99} C.A. Bertulani and V. Ponomarev, Phys. Reports 321 (1999).
\bibitem{Bert96} C.A. Bertulani, L.F. Canto, M.S. Hussein and A.F.R. de Toledo Piza, Phys. Rev. C 53, 334 (1996). 
\bibitem{Ber05} C. A. Bertulani, Phys. Rev. Lett. 94, 072701 (2005).
\bibitem{BC10} C.A. Bertulani and C. De Conti, Phys. Rev. C 81, 064603 (2010).
\bibitem{Kar13} Mesut Karakoc, A. Banu, C.A. Bertulani, L. Trache, Phys. Rev. C 87, 024607 (2013).
\bibitem{Vri87} H. de Vries, C.W. de Jager, C. de Vries, At. Nucl. Data Tables 36, 495, 536 (1987).
\bibitem{ABS95} T. Aumann, C.A. Bertulani and K. Suemmerer, Phys. Rev. C 51, 416 (1995). 
\bibitem{Wie09} O. Wieland et al., Phys. Rev. Lett. 102, 092502 (2009).
\bibitem{Ros13} D. M. Rossi at al., Phys. Rev. Lett. 111, 242503 (2013).
\bibitem{Sav13} D. Savran, T. Aumann und, A. Zilges, Prog. Part. Nucl. Phys. 70, 210 (2013).
\bibitem{Youn92} D.H. Youngblood, Y.-W. Lui, U. Garg and R.J. Peterson, Phys. Rev. C 45, 2172 (1992).
\bibitem{Bri55} D.M. Brink, Ph.D. thesis, Oxford University, 1955.
\bibitem{Axe62} P. Axel, Phys. Rev. 126, 671 (1962).
\bibitem{Pie11} J. Piekarewicz, Phys. Rev. C 83, 034319 (2011).
\bibitem{PB98} V.Yu. Ponomarev and C.A. Bertulani, Phys. Rev. C 57 (1998) 3476.


\end{thebibliography}
\end{document}